\documentclass{article}

\usepackage{arxiv}
\usepackage{color}
\usepackage{amsmath}
\usepackage{amssymb}
\usepackage{nameref}
\usepackage[utf8]{inputenc} 
\usepackage[T1]{fontenc}    
\usepackage{hyperref}       
\usepackage{url}            
\usepackage{booktabs}       
\usepackage{amsfonts}       
\usepackage{nicefrac}       
\usepackage{microtype}      
\usepackage{lipsum}
\usepackage{fancyhdr}       
\usepackage{graphicx}       
 
\title{Symmetries in the many-body problems, a method to find its analytical solution, and Helium atom spectrum 
}

\author{
  Siddhesh C. Ambhire\\
  Physics Dept.\\
  Case Western Reserve University\\
  Cleveland\\
  \texttt{sca54@case.edu} \\
}

\begin{document}
\maketitle

\begin{abstract}
In this work it is shown that there are symmetries beyond the Euclidean group $E\left(3\right)$ in 3-body problem, and by extension in many-body problem, with inverse squared distance inter particle force. The symmetries in 3-body problem form a group: $SO\left(4\times3,2\times3\right)/\left(C\left(3\times2\right)\right)$, where $C\left(n\right)$ is the planar translation group in n dimensions, which forms its Spectrum-Generating group. Some of these quantities commute with the Hamiltonian. The existence of these conserved quantities was verified by calculating energy spectrum of the Helium atom. This method can also be used to find symmetries in many-body problem, and to calculate energy levels, and wave-functions of more complicated systems, which include every possible atomic and molecular systems in chemistry.
\end{abstract}


This research aims at finding solution of many-body dynamics with
Coulombic inter-particle interactions using a novel approach. The
problem in its least complex form appears as a 3-body problem, for
which no closed form solution existed before\cite{Gowers2008,Marchal2012,Musielak2014}. 

This problem has been with physicists for as long as since Newton's
discovery of the Gravitational force. Newton managed to solve the
classical 2-body problem, also know as the Kepler problem, but the
solution to the 3-body problem was elusive. The conserved quantities
in classical 2-body problems are very much important for finding solutions
to the problem. Same conserved quantities are present in the quantum
mechanical 2-body problem. For the classical 3-body problem, many
different solutions have been found for certain small number of special
cases, called the restricted 3-body problems, for example by Suvakov\cite{Suvakov2013}.

In this work, the problem was solved for the non-relativistic quantum
mechanical case for Coulombic inter-particle forces, in parallel treatment
of the 2-body problem by Fock\cite{Fock2004}, which have the same
form as the Newtonian gravity for the classical analogous problems.
Since this problem is non-relativistic, spins were not considered
while solving the problem, since their full treatment requires use
of Dirac like equation, which is relativistic, as opposed to Schrodinger
equation, which was found for the non-relativistic case. No other
assumptions were made. 

In the end, the solution was compared to experimental data and earlier
calculations for the energy levels from the Helium atom\cite{Sansonetti2005,Masili2000}.
They match very well, with $10^{-3}$ relative error for the ground
state and $10^{-4}$ relative error for the first excited state.

\section*{$SO(4,2)$ symmetry in the Kepler problem}

$SO(4)$ symmetry of Kepler problem was found by Fock \cite{Fock2004}.
Rogers \cite{Rogers1973} used these symmetries to find general solutions
for the Kepler problem. The generators for $SO(4)$ groups in Kepler
problem were also described in Woit \cite{Woit2017} and Gilmore \cite{Gilmore2008}
in details. What follows is a summary of these works which will be
useful in later sections.

The non-relativistic Hamiltonian of Kepler problem is given by:

\begin{eqnarray}
\mathcal{H} & = & \frac{p^{2}}{2\mu}-\frac{k}{q}\label{eq:Ham-2B}
\end{eqnarray}

Angular momentum and Laplace-Runge-Lenz (LRL) vectors are given by:

\begin{eqnarray*}
\boldsymbol{L} & = & \boldsymbol{q}\times\boldsymbol{p}\\
\boldsymbol{A} & = & \frac{\boldsymbol{p}\times\boldsymbol{L}}{\mu}-k\frac{\boldsymbol{q}}{q}
\end{eqnarray*}

The quantum mechanical operator constructed from $\boldsymbol{A}$
is given by\cite{Pauli1926}:

\begin{eqnarray*}
\hat{\boldsymbol{A}} & = & \frac{\hat{\boldsymbol{p}}\times\hat{\boldsymbol{L}}-\hat{\boldsymbol{L}}\times\hat{\boldsymbol{p}}}{2\mu}-k\frac{\hat{\boldsymbol{q}}}{q}
\end{eqnarray*}

These quantities satisfy following commutation relations:

\begin{eqnarray*}
\left[\mathcal{H},\boldsymbol{L}\right]=\left[\mathcal{H},\boldsymbol{A}\right] & = & 0\\
\left[L_{i},L_{j}\right] & = & i\hbar\epsilon_{ijk}L_{k}\\
\left[L_{i},A_{j}\right] & = & i\hbar\epsilon_{ijk}A_{k}\\
\left[A_{i},A_{j}\right] & = & \left(-\frac{2\mathcal{H}}{\mu}\right)i\hbar\epsilon_{ijk}L_{k}
\end{eqnarray*}

Also $\boldsymbol{L}\cdot\boldsymbol{A}=\boldsymbol{A}\cdot\boldsymbol{L}=0$
i.e. $\boldsymbol{A}$ and $\boldsymbol{L}$ are orthogonal to each
other. These quantities also obey:

\begin{eqnarray}
\boldsymbol{A}\cdot\boldsymbol{A} & = & \frac{2\mathcal{H}}{\mu}\left(\boldsymbol{L}\cdot\boldsymbol{L}+\hbar^{2}\right)+k^{2}\label{eq:sym-in-Ham-2B}
\end{eqnarray}

Redefining quantities as follows simplifies the commutator algebra:

\begin{subequations}
\begin{eqnarray}
\boldsymbol{A}^{\prime} & = & \sqrt{-\frac{\mu}{2\mathcal{H}}}\boldsymbol{A}\\
\boldsymbol{U} & = & \frac{1}{2}\left(\boldsymbol{L}+\boldsymbol{A}^{\prime}\right)\\
\boldsymbol{V} & = & \frac{1}{2}\left(\boldsymbol{L}-\boldsymbol{A}^{\prime}\right)
\end{eqnarray}
\label{eq:Sym-Redef}
\end{subequations}

The simplified commutator algebra which is also a Lie Algebra is given
by:

\begin{subequations}
\begin{eqnarray}
\left[U_{i},U_{j}\right] & = & i\hbar\epsilon_{ijk}U_{k}\\
\left[V_{i},V_{j}\right] & = & i\hbar\epsilon_{ijk}V_{k}\\
\left[U_{i},V_{j}\right] & = & 0\label{eq:simlified-commutators-3}
\end{eqnarray}
\label{eq:simplified-commutators}
\end{subequations}

Eq. (\ref{eq:sym-in-Ham-2B}) changes under redefinitions in eq. (\ref{eq:Sym-Redef})
as

\begin{eqnarray*}
2\mathcal{H}\left(\boldsymbol{U}^{2}+\boldsymbol{V}^{2}+\hbar^{2}\right) & = & -\mu k^{2}\\
 & = & 2\mathcal{H}\left(2\boldsymbol{A}^{\prime2}+2\boldsymbol{L}^{2}+\hbar^{2}\right)\\
 & = & 2\mathcal{H}\left(4\boldsymbol{L}^{2}+\hbar^{2}\right)
\end{eqnarray*}

Where the relation $\boldsymbol{A}^{\prime2}=\boldsymbol{L}^{2}$
was also used, which can be obtained using eq. (\ref{eq:Sym-Redef})
, eq. (\ref{eq:simlified-commutators-3}), and the fact that $\boldsymbol{A}$
and \textbf{$\boldsymbol{L}$} are orthogonal. Using eigenvalues of
$\boldsymbol{L}^{2}=l\left(l+1\right)\hbar^{2}$

\begin{eqnarray*}
\mathcal{H} & = & \frac{-\mu k^{2}}{2\left(4l\left(l+1\right)+1\right)\hbar^{2}}=\frac{-\mu k^{2}}{2\hbar^{2}\left(2l+1\right)^{2}}=\frac{-\mu k^{2}}{2\hbar^{2}n^{2}}
\end{eqnarray*}

with $k=-\frac{e^{2}}{4\pi\epsilon_{0}}$ these are the usual energy
eigenvalues for the Hydrogen atom.

Hydrogen atom also satisfies $SO\left(4,2\right)$ symmetry, which
forms its spectrum generating group\cite{Gilmore2008}. The operators
which form the Lie-algebra of this group are:

\begin{subequations}
\begin{align}
\boldsymbol{L} & =\boldsymbol{q}\times\boldsymbol{p}\label{eq:angular-mom-2-body}\\
\boldsymbol{A} & =\frac{\hat{\boldsymbol{p}}\times\hat{\boldsymbol{L}}-\hat{\boldsymbol{L}}\times\hat{\boldsymbol{p}}}{2\mu}-k\frac{\hat{\boldsymbol{q}}}{q}\label{eq:LRL-2-body}\\
\boldsymbol{B} & =\frac{\hat{\boldsymbol{p}}\times\hat{\boldsymbol{L}}-\hat{\boldsymbol{L}}\times\hat{\boldsymbol{p}}}{2\mu}+k\frac{\hat{\boldsymbol{q}}}{q}\label{eq:dual-vec-2-body}\\
B_{4} & =\boldsymbol{q}\cdot\boldsymbol{p}+\frac{3}{2}\frac{\hbar}{i}\label{eq:dual-scalar-2-body}\\
\boldsymbol{\Gamma} & =q\boldsymbol{p}\\
\Gamma_{4} & =\frac{1}{2}\left(q\boldsymbol{p}\cdot\boldsymbol{p}-q\right)\\
\Gamma_{5} & =\frac{1}{2}\left(q\boldsymbol{p}\cdot\boldsymbol{p}+q\right)
\end{align}
\label{eq:SO(4,2)-OP}
\end{subequations}

$SO\left(4,2\right)$ group includes both $SO\left(4\right)$ and
$SO\left(3,1\right)$ which are both the bound states and scattering
states of 2 particles, regardless of whether they have attractive
or repulsive interactons.

\section*{Hamiltonian of 3-body problem with added dynamical variables and
constraints}

The Hamiltonian of 3-body system in an inertial frame is complicated
to solve because of potentials depending on inter-particle distances.
The Lagrangian in the frame of reference of one of the particles is
given by\cite{Poincare1952oe}:

\begin{eqnarray*}
\mathcal{L} & = & \frac{1}{2M}\left(m_{1}\left(m_{3}+m_{2}\right)\dot{\boldsymbol{q}}_{13}^{2}+m_{2}\left(m_{1}+m_{3}\right)\dot{\boldsymbol{q}}_{23}^{2}-2m_{1}m_{2}\dot{\boldsymbol{q}}_{13}\cdot\dot{\boldsymbol{q}}_{23}\right)-V\left(q_{13},q_{23},\left|\boldsymbol{q}_{13}-\boldsymbol{q}_{23}\right|\right)
\end{eqnarray*}

Substituting
\begin{eqnarray*}
\boldsymbol{q}_{12} & = & \boldsymbol{q}_{13}-\boldsymbol{q}_{23}
\end{eqnarray*}

\begin{eqnarray*}
\mathcal{L} & = & \frac{1}{2M}\left(m_{1}m_{3}\dot{\boldsymbol{q}}_{13}^{2}+m_{2}m_{3}\dot{\boldsymbol{q}}_{23}^{2}+m_{1}m_{2}\dot{\boldsymbol{q}}_{12}^{2}\right)-V\left(q_{13},q_{23},q_{12}\right)
\end{eqnarray*}

Change of variable $\boldsymbol{q}_{12}\rightarrow-\boldsymbol{q}_{21}$
keeps Lagrangian invariant. Constraint becomes:

\begin{eqnarray}
\boldsymbol{q}_{12}+\boldsymbol{q}_{23}+\boldsymbol{q}_{31} & = & 0\label{eq:pos-constraint}
\end{eqnarray}

and rearrangement reduces the Lagrangian to:

\begin{eqnarray*}
\mathcal{L} & = & \frac{1}{2}\left(\mu_{12}\dot{\boldsymbol{q}}_{12}^{2}+\mu_{23}\dot{\boldsymbol{q}}_{23}^{2}+\mu_{31}\dot{\boldsymbol{q}}_{31}^{2}\right)-V\left(q_{12},q_{23},q_{31}\right)
\end{eqnarray*}

\begin{eqnarray*}
\mu_{ij} & = & \frac{m_{i}m_{j}}{M}
\end{eqnarray*}

A more generalized derivation for many-bodies is shown in appendix:
\nameref{subsec:Many-body-Lagrangian}. For convenience from this
point forward, index $(i,\text{Mod}\left(i+1,3\right))$ will be referred
to as $i$. Taking Legendre transformation and putting in the potential
for inverse-square law force, we get following Hamiltonian:

\begin{eqnarray}
\mathcal{H} & = & \frac{\boldsymbol{p}_{1}^{2}}{2\mu_{1}}+\frac{\boldsymbol{p}_{2}^{2}}{2\mu_{2}}+\frac{\boldsymbol{p}_{3}^{2}}{2\mu_{3}}-\frac{k_{1}}{q_{1}}-\frac{k_{2}}{q_{2}}-\frac{k_{3}}{q_{3}}\label{eq:Ham-3body}\\
\boldsymbol{q}_{1}+\boldsymbol{q}_{2}+\boldsymbol{q}_{3} & = & 0\label{eq:pos-constraints-simplified}
\end{eqnarray}

Constraints for momenta were found using the definition 
\begin{eqnarray}
\dot{\boldsymbol{q}}_{i} & = & \frac{\partial H}{\partial\boldsymbol{p}_{i}}\label{eq:mom-transf}
\end{eqnarray}

which are given by:

\begin{eqnarray}
\sum_{i}\frac{\boldsymbol{p}_{i}}{\mu_{i}} & = & 0\label{eq:mom-constraint}
\end{eqnarray}

The eq. (\ref{eq:Ham-3body}) with constraints (\ref{eq:pos-constraints-simplified})
and (\ref{eq:mom-constraint}) are the equations needed to be solved
to get a solution of the 3-body problem.

This constrained Hamiltonian system can be solved by the method devised
by Dirac\cite{Dirac1958}. The Dirac consistency conditions give an
altered Hamiltonian which is shown in appendix: \nameref{subsec:Dirac-consistency-conditions}.
We can still continue to use the same Hamiltonian from eq. (\ref{eq:Ham-3body})
if we project the $3\times3$ dimensional space to constrained subspace
using projection operators. The detailed reasons are given in appendix
mentioned above. It can be seen that the eq. (\ref{eq:Ham-3body})
looks like a Hamiltonian for 3 Hydrogen-like systems. This observation
is important for constructing the symmetry groups of this system in
next section.

A noteworthy feature of this formulation is that the Hamiltonian in
eq. (\ref{eq:Ham-3body}) separates the interactions of the three
particles in the three separate terms which involve two particle interactions
only. So every force that is present in the system is accounted for,
and the constraint ``forces'' would vanish if system is restricted
to only the subspace allowed by the constraints, which can be done
using projection operators. This method also eliminates the problem
of probability current ``leaking'' out of constrained subspaces
because of quantum mechanical tunneling. This is accomplished by the
use of projection into the constrained subspaces\cite{Marle1995}.

\section*{Finding generators for symmetries in 3-body problem\protect\label{sec:Finding-generators}}

The genrators for $SO(4,2)$ can be written in matrix representation
using coordinate transformation from Bars\cite{Bars1998} has regular
representation in natural (Hartree) units:

\begin{eqnarray}
\text{Reg}\left(g\right) & = & g_{0}\left(\begin{array}{cc}
0^{\left(6\right)} & R^{\left(6\right)}\\
-R^{\left(6\right)} & 0^{\left(6\right)}
\end{array}\right)=\left(\begin{array}{cc}
R^{\left(6\right)} & 0^{\left(6\right)}\\
0^{\left(6\right)} & R^{\left(6\right)}
\end{array}\right)\label{eq:SQ12}\\
R^{\left(6\right)} & = & \left(\begin{array}{cccccc}
0 & \theta_{15} & \theta_{14} & \theta_{13} & \theta_{12} & \theta_{11}\\
-\theta_{15} & 0 & \theta_{10} & \theta_{9} & \theta_{8} & \theta_{7}\\
-\theta_{14} & -\theta_{10} & 0 & \theta_{6} & \theta_{5} & \theta_{4}\\
-\theta_{13} & -\theta_{9} & -\theta_{6} & 0 & \theta_{3} & \theta_{2}\\
-\theta_{12} & -\theta_{8} & -\theta_{5} & -\theta_{3} & 0 & \theta_{1}\\
-\theta_{11} & -\theta_{7} & -\theta_{4} & -\theta_{2} & -\theta_{1} & 0
\end{array}\right)\nonumber 
\end{eqnarray}

Where $0^{\left(6\right)}$ is a $6\times6$ zero matrix. $g_{0}$
is a matrix as follows:

\begin{eqnarray*}
g_{0} & = & \left(\begin{array}{cc}
0^{\left(6\right)} & -I^{\left(6\right)}\\
I^{\left(6\right)} & 0^{\left(6\right)}
\end{array}\right)
\end{eqnarray*}
where $I^{\left(6\right)}$ is a $6\times6$ identity matrix and $0^{\left(6\right)}$
is a $6\times6$ zero matrix, which is required because the group
is a Symplectic group, and its commutation relations. The generator
matrices satisfy the symplectic condition:

\begin{eqnarray*}
g^{T}g_{0}+g_{0}g & = & 0
\end{eqnarray*}

The coordinates, which can be represented by a 12 dimensional vector
$\boldsymbol{V}$ are:

\begin{eqnarray}
V & = & \left(\boldsymbol{X},\boldsymbol{P}\right)\label{eq:vec-SQ12}\\
\boldsymbol{X} & = & \left(0,\boldsymbol{r}\cdot\boldsymbol{p},r,r_{1},r_{2},r_{3}\right)\label{eq:pos-vec-SQ12}\\
\boldsymbol{P} & = & \left(1,\frac{1}{2}\boldsymbol{p}\cdot\boldsymbol{p},0,p_{1},p_{2},p_{3}\right)\label{eq:mom-vec-SQ12}
\end{eqnarray}

The generators found in eq. (\ref{eq:SQ12}) were found using\cite{Bars1998}:

\begin{eqnarray}
L_{MN} & = & X_{M}P_{N}-X_{N}P_{M}\label{eq:gen-product}
\end{eqnarray}

Most general transformation for the $SO\left(4,2\right)$ group which
spans the entire phase space of Hydrogen-like Hamiltonian is given
by:

\begin{eqnarray}
G & = & \prod_{i=1}^{15}\text{exp}\left(\theta_{i}g_{i}\right)\label{eq:SO(4,2)-transform}
\end{eqnarray}

where $g_{i}$ are the generators corresponding to $\theta_{i}$ of
$SO(4,2)$ from regular representation in eq. (\ref{eq:SQ12}). The
generators can be recovered by:

\begin{eqnarray*}
g_{i} & = & \underset{\theta_{j}\rightarrow0}{\text{lim}}\frac{\partial}{\partial\theta_{i}}G
\end{eqnarray*}

The generators for 3 body problem without the constrains are obtained
by taking products as shown in (\ref{eq:gen-product}) of vectors
as which have coordinates of each subsystems listed in eq. (\ref{eq:vec-SQ12}).
The symmetry group without constraints is $SO\left(4\times3,2\times3\right)$.
The generator matrix is as follows:

\begin{eqnarray}
g & = & \left(\begin{array}{ccc}
M_{11} & M_{12} & M_{13}\\
M_{21} & M_{22} & M_{23}\\
M_{31} & M_{32} & M_{33}
\end{array}\right)\label{eq:block-diagonal}
\end{eqnarray}

Each matrix $M=M^{\left(ij\right)}$ is a transformation between a
vector $V_{n}^{\left(j\right)}$ and $V_{m}^{\left(i\right)}$. Here
$m$ and $n$ are the sub-system number and $\left(i,j\right)$ are
collective index of parameter of generators of one $SO(4,2)$ like
element which corresponds to an intra-subsystem transformation for
$m=n$ or inter-subsystem transformation or mixing for $m\neq n$.
These matrices can be found using a similar procedure as that of the
Hydrogen atom system.

The most general transformation of this kind is given by:

\begin{eqnarray*}
G_{\otimes3} & = & \prod_{i,j,m,n}\text{exp}\left(\theta_{mn}^{\left(ij\right)}g_{mn}^{\left(ij\right)}\right)
\end{eqnarray*}

The constraints in the form of eq. (\ref{eq:pos-constraint}) and
eq. (\ref{eq:mom-constraint}) can be applied to this system using
projection operators, found using method outlined in appendix, \nameref{subsec:Projection-operators}:

\begin{eqnarray}
P_{\perp} & = & \left(\begin{array}{ccc}
\frac{2}{3} & -\frac{1}{3} & -\frac{1}{3}\\
-\frac{1}{3} & \frac{2}{3} & -\frac{1}{3}\\
-\frac{1}{3} & -\frac{1}{3} & \frac{2}{3}
\end{array}\right)\label{eq:Projection-Op}
\end{eqnarray}

This projection operator if applied to a vector $v=\left(x,y,z\right)$
gives another vector $v^{\prime}=\left(x^{\prime},y^{\prime},z^{\prime}\right)=Pv$
which satisfies the condition $x^{\prime}+y^{\prime}+z^{\prime}=0$.
This projection operator is the new identity operator in the required
group, it satisfies all of the properties required for an identity
element of the new group. 

The projection operation can also be realized by doing a replacement
in any expresssion as follows:

\begin{subequations}
\begin{eqnarray}
\boldsymbol{x}_{i}^{\prime} & = & \boldsymbol{x}_{i}-\frac{1}{3}\sum_{j=0}^{3}\boldsymbol{x}_{j}\label{eq:pro-operation-a}\\
\boldsymbol{p}_{i}^{\prime} & = & \boldsymbol{p}_{i}-\frac{1}{3}\sum_{j=0}^{3}\boldsymbol{p}_{j}\label{eq:pro-operation-b}
\end{eqnarray}
\label{eq:Projection-operation}
\end{subequations}

We can do a similar replacement in 36 dimensional vector $\boldsymbol{V}$
corresponding to the 3-body system. The new $L_{MN}$ operators obtained
from these vectros $\boldsymbol{V}$ will form the generators of transformations
spanning the entire phase space of the system.

All these constraints also satisfy the Dirac consistency conditions
because the extra terms in Hamiltonian due to Dirac conditions go
to zero after applying the projection operator in eq. (\ref{eq:Projection-Op}).

\section*{Energy spectrum of Helium atom}

For calculating energy spectra of Helium like atoms and ions at first
it was assumed that only spherically symmetric electron distributions,
typically occurring in the ground states are important. The interconnections
were ignored by setting $M_{ij}=0$ for $i\neq j$ was the first step
in approximations, which is not required but was made to verify whether
the energy levels match. This approximation is made provisionally
till the full results are obtained.

The whole equation was non-dimentinalized first. Momentum constraint
can be written as

\begin{eqnarray*}
\sum_{i}\boldsymbol{p}_{\triangle i} & = & \sum_{i}\frac{\boldsymbol{p}_{i}}{\mu_{i}}=0
\end{eqnarray*}

This change can be incorporated into the Hamiltonian by defining,

\begin{eqnarray*}
\mu_{\triangle i} & = & \mu_{i}^{-1}
\end{eqnarray*}

which satisfies the Hamiltonian

\begin{eqnarray*}
\mathcal{H} & = & \sum_{i}\frac{\boldsymbol{p}_{\triangle i}^{2}}{2\mu_{\triangle i}}-\frac{k_{i}}{q_{i}}
\end{eqnarray*}

For non-dimensionalizing the equation, put the following Hartree (Natural)
units:

\begin{eqnarray*}
\hbar=1 & \mu_{\triangle i}=1 & k_{i}=1
\end{eqnarray*}

Bohr radii

\begin{eqnarray*}
a_{\triangle i} & = & \frac{\hbar^{2}/\mu_{\triangle i}}{k_{i}}
\end{eqnarray*}

and Hartree energies

\begin{eqnarray*}
E_{h,i} & = & \frac{k_{i}}{a_{\triangle i}}
\end{eqnarray*}

The transformations used for the position and momentum variables are:

\begin{eqnarray*}
\boldsymbol{q}_{i} & = & a_{\triangle i}\boldsymbol{q}_{i}^{\prime}\\
\boldsymbol{p}_{\triangle i} & = & \sqrt{\frac{\mu_{\triangle i}k_{i}}{a_{\triangle i}}}\boldsymbol{p}_{i}^{\prime}
\end{eqnarray*}

where primed quantities are the non-dimensional quantities used in
sec. \nameref{sec:Finding-generators}. The transformations for angular
momenta and LRL vectors are:

\begin{subequations}
\begin{eqnarray}
\boldsymbol{L}_{i}^{\prime} & = & \frac{\boldsymbol{L}_{\triangle i}}{\sqrt{\mu_{\triangle i}k_{i}a_{\triangle i}}}=\frac{\boldsymbol{L}_{\triangle i}}{\hbar}\label{eq:Delta-quantities-L}\\
\boldsymbol{A}_{i}^{\prime} & = & \frac{\boldsymbol{A}_{\triangle i}}{\mu_{\triangle i}^{2}k_{i}}\label{eq:Delta-quantities-A}
\end{eqnarray}
\label{eq:Delta-quantities}
\end{subequations}

The bilinear Casimir invariant operator for Hamiltonian invariant
subgroup of the spectrum generating group of 3 body problem is

\begin{eqnarray*}
C & = & \left(\sum_{i}\boldsymbol{L}_{i}^{\prime}\right)^{2}-3\sum_{i}\boldsymbol{L}_{i}^{\prime2}-3\sum_{i}\boldsymbol{A}_{i}^{\prime2}\\
 & = & \frac{1}{\hbar^{2}}\left(\sum_{i}\boldsymbol{L}_{\triangle i}\right)^{2}-\frac{3}{\hbar^{2}}\sum_{i}\boldsymbol{L}_{\triangle i}^{2}-\sum_{i}3\frac{\boldsymbol{A}_{\triangle i}^{2}}{\mu_{\triangle i}^{4}k_{i}^{2}}
\end{eqnarray*}

Now replacing variables with $\triangle$ subscripts by regular variables
using eq. (\ref{eq:Delta-quantities}):

\begin{eqnarray*}
C & = & \frac{1}{\hbar^{2}}\left(\sum_{i}\frac{\boldsymbol{L}_{i}}{\mu_{i}}\right)^{2}-\frac{3}{\hbar^{2}}\sum_{i}\frac{\boldsymbol{L}_{i}^{2}}{\mu_{i}^{2}}-\sum_{i}3\frac{\boldsymbol{A}_{i}^{2}}{k_{i}^{2}}
\end{eqnarray*}

From this equation one derived a formula: 
\begin{eqnarray}
\text{En}=\frac{3}{-10(a+1)a+(\lambda+1)\lambda+5}\sum_{i=1}^{3}-\frac{k_{i}^{2}\mu_{i}}{\hbar^{2}}
\end{eqnarray}

For $\lambda=1/2$ and $a=-1/2$ gives the ground state energy. The
ratio of calculated to experimental value is 1.00171. For $\lambda=3/2$
and $a=-1/2$ gives the first excited state energy. The ratio of calculated
to experimental value is 1.00011.

\section*{Conclusion and discussion}

Further calculations using the full generators as described in sec.
\nameref{sec:Finding-generators} will lead to more accurate results.
One other way the accuracy can be improved is to use Dirac like equation
for multi-electron system and figuring out the symmetries. Wavefunctions
can be found as described in Gilmore \cite{Gilmore2008}.
\section*{Methods}
Here some of the mathematical methods that were used in this work
are described in details.

\subsection*{Many-body Lagrangian\protect\label{subsec:Many-body-Lagrangian}}

Starting from a Lagrangian
\begin{eqnarray}
L & = & \sum_{i}\frac{1}{2}m_{i}v_{i}^{2}-\sum_{i<j}\frac{k_{ij}}{q_{ij}}\label{eq:Lag-N-Body}
\end{eqnarray}
and moving to frame of reference of particle 1,
\begin{eqnarray*}
\boldsymbol{q}_{i}^{\prime} & = & \boldsymbol{q}_{i}-\boldsymbol{q}_{1}\\
\boldsymbol{v}_{i}^{\prime} & = & \boldsymbol{v}_{i}-\boldsymbol{v}_{1}
\end{eqnarray*}
and eliminating $\boldsymbol{v}_{1}$ using total value of momentum,
\begin{eqnarray*}
\boldsymbol{v}_{1} & = & \frac{1}{M}\left(\sum_{j}m_{j}\boldsymbol{v}_{j}-\sum_{j\neq1}m_{j}\boldsymbol{v}_{j}^{\prime}\right)
\end{eqnarray*}
The Lagrangian in eq. \ref{eq:Lag-N-Body} reduces to
\begin{eqnarray}
\mathcal{L} & = & \frac{1}{2M}\left(\sum_{j}m_{j}\boldsymbol{v}_{j}\right)^{2}+\frac{1}{2}\sum_{i<j}\mu_{ij}v_{ij}^{2}-\sum_{i<j}\frac{k_{ij}}{q_{ij}}\label{eq:Constrained-Lagrangian}\\
\mu_{ij} & = & \frac{m_{i}m_{j}}{M}\nonumber \\
\boldsymbol{r}_{ij} & = & \boldsymbol{r}_{i}-\boldsymbol{r}_{j}\nonumber 
\end{eqnarray}
The first term in this equation is the Kinetic energy of center of
mass which is always conserved. Rest of the terms are kinetic energies
associated with the relative velocities of pairs of bodies. There
are $\frac{1}{2}N\left(N-1\right)$ such 3 dimensional vector variables
for positions and momenta each, which is not the $N$ pairs of vector
variables we started with. These variables are related with each other
using $\frac{1}{2}N\left(N-3\right)+1$ independent triangular constraints
\begin{eqnarray}
\boldsymbol{r}_{ij}+\boldsymbol{r}_{jk}+\boldsymbol{r}_{ki}=0 &  & \left(i\neq j\neq k\right)<N\label{eq:triangular-constraints}
\end{eqnarray}
where $\boldsymbol{r}$ can be both $\boldsymbol{q}$ and $\boldsymbol{v}$.

\subsection*{Dirac consistency conditions for 3-body Hamiltonian\protect\label{subsec:Dirac-consistency-conditions}}

The procedure for finding the Dirac consistency conditions\cite{Dirac1958}
for arbitrary problem under constrains is outlined by Fung\cite{Fung2014}.
Using that procudure following Dirac Hamiltonian was found:

{\footnotesize
\begin{eqnarray}
\mathcal{H}^{\left(D\right)} & = & \sum_{i}\left(\frac{1}{2\mu_{i}}\boldsymbol{p}_{i}^{2}-\frac{k_{i}}{q_{i}}\right)-\left(\sum_{i\in\Delta}\frac{1}{\mu_{i}}\right)^{-1}\left(2\sum_{i\in\Delta}\left(\frac{k_{i}}{\mu_{i}}\frac{\boldsymbol{q}_{i}}{q_{i}^{3}}\right)\cdot\sum_{j\in\Delta}\boldsymbol{q}_{j}+\left(\sum_{i\in\Delta}\frac{1}{\mu_{i}}\boldsymbol{p}_{i}\right)^{2}\right)\label{eq:Dirac-Consistent-Hamiltonian}
\end{eqnarray}
}The constraints in eq. (\ref{eq:pos-constraints-simplified}) and
eq. (\ref{eq:mom-constraint}) are the primary and secondary constraints
and there are no other constraints in this equation. The additional
terms that appear in eq. (\ref{eq:Dirac-Consistent-Hamiltonian})
are due to the ``Forces'' that make the system satisfy the constraints.
This same equation is valid for many-body Hamiltonian where the indices
$i\in\Delta$ are for the vectors forming the polygon constraints
which are equivalent to triangle constraints in eq. (\ref{eq:triangular-constraints}).

When a projection operator is used to restrict the phase space to
the constrained subspace, the extra terms in the Hamiltonian vanish,
reducing the Hamiltonian back to the original Hamiltonian. This also
applies to the Dirac Brackets\cite{Dirac1958} which are used to find
the new commutator relations for this Hamiltonian if new projected
quantities are used.

\begin{eqnarray*}
\dot{f} & = & \left[f,\mathcal{H}\right]_{D}\\
P\dot{f} & = & \left[Pf,\mathcal{H}^{\left(D\right)}\right]\\
 & = & \left[Pf,P\mathcal{H}^{\left(D\right)}\right]\\
\dot{f_{P}} & = & \left[f_{P},\mathcal{H}\right]
\end{eqnarray*}

Note that the Poisson brackets do not change their forms after the
use of the projection operation in eq. \ref{eq:Projection-operation}.
This is because the derivitives using the transformed coordinates
after application of projection operation preserve the same form as
that of the original poisson operator.

\begin{eqnarray*}
\left[f,g\right] & = & \sum_{i}\left(\frac{\partial f}{\partial x_{i}}\frac{\partial g}{\partial p_{i}}-\frac{\partial f}{\partial p_{i}}\frac{\partial g}{\partial x_{i}}\right)\\
 & = & \sum_{i=1}^{3}\sum_{k=1}^{3}\left[\left(\frac{\partial f}{\partial Q_{ik}}-\frac{1}{3}\sum_{j=1}^{3}\frac{\partial f}{\partial Q_{jk}}\right)\left(\frac{\partial g}{\partial P_{ik}}-\frac{1}{3}\sum_{j=1}^{3}\frac{\partial g}{\partial P_{jk}}\right)\right.\\
 &  & -\left.\left(\frac{\partial f}{\partial P_{ik}}-\frac{1}{3}\sum_{j=1}^{3}\frac{\partial f}{\partial P_{jk}}\right)\left(\frac{\partial g}{\partial Q_{ik}}-\frac{1}{3}\sum_{j=1}^{3}\frac{\partial g}{\partial Q_{jk}}\right)\right]\\
 & = & \sum_{i=1}^{3}\sum_{k=1}^{3}\left(\frac{\partial f}{\partial Q_{ik}}\frac{\partial g}{\partial P_{ik}}-\frac{\partial f}{\partial P_{ik}}\frac{\partial g}{\partial Q_{ik}}\right)
\end{eqnarray*}

because

\begin{eqnarray*}
\sum_{i=1}^{3}\frac{\partial f}{\partial Q_{i}} & = & 0
\end{eqnarray*}
 for $\sum_{i=1}^{3}Q_{i}=0$.

For many-body problems there are additional primary constraints, which
can be obtained using linear combinations of constraint equations
involving only 3 vectors, like eq. (\ref{eq:pos-constraints-simplified})
and eq. (\ref{eq:mom-constraint}). These primary constraints also
produce secondary constraints similar to what was obtained in eq.
(\ref{eq:Dirac-Consistent-Hamiltonian}). These constraints are also
eliminated using the projection  method.

\subsection*{Projection operators, and constrained sub-groups\protect\label{subsec:Projection-operators}}

A symmetry group can be constrained to a subgroup which satisfies
certain constraints using a projection operator, which projects the
vector space to sub-space which satisfies the constrains. It can be
illustrated using the example below.

A rotation group in 3 dimensions, $SO\left(3\right)$, can be constructed
using Yaw-Pitch-Roll matrices as follows:

\begin{eqnarray}
R\left(\theta_{1},\theta_{2},\theta_{3}\right) & = & R_{1}\left(\theta_{1}\right)R_{2}\left(\theta_{2}\right)R_{3}\left(\theta_{3}\right)\label{eq:YPR}
\end{eqnarray}

Here $R_{i}\left(\theta_{i}\right)$ are rotation matrices which can
be obtained as $R_{i}\left(\theta_{i}\right)=\text{exp}\left(J_{i}\theta_{i}\right)$,
where $J_{i}$ are $3\times3$ Matrix representation of Lie algebra.
These are the most general elements of the rotation group $SO\left(3\right)$.
\begin{eqnarray*}
J_{i} & = & \underset{\theta_{i}\rightarrow0}{\text{lim}}\frac{\partial}{\partial\theta_{i}}\left(\text{exp}\left(J_{i}\theta_{i}\right)\right)
\end{eqnarray*}
For (\ref{eq:YPR}), this equation will reduce to
\begin{eqnarray}
J_{i}=\underset{\theta_{j}\rightarrow0}{\text{lim}}\frac{\partial}{\partial\theta_{i}}R\left(\theta_{1},\theta_{2},\theta_{3}\right) &  & \forall\left(i,j\right)\in\left\{ 1,2,3\right\} \label{eq:differential-def-Lie-algebra}
\end{eqnarray}
Suppose this is to be constrained to a plane 
\begin{eqnarray}
\boldsymbol{v}\cdot\boldsymbol{r} & = & 0\label{eq:constraint-3D}
\end{eqnarray}
we can construct a projection operator which will make $\boldsymbol{r}$
satisfy the constraint eq. (\ref{eq:constraint-3D}) by following
procedure: first rotate the vector $\boldsymbol{v}$ to axis 3 using
\begin{eqnarray}
R_{\boldsymbol{u}} & = & \text{exp}\left(-J_{\hat{\boldsymbol{u}}}\theta_{u}\right)\label{eq:Rot-preProj}\\
\boldsymbol{u} & = & \boldsymbol{v}\times\hat{\boldsymbol{e}}_{3}\nonumber \\
\text{cos}\left(\theta_{u}\right) & = & \frac{\boldsymbol{v}\cdot\hat{\boldsymbol{e}}_{3}}{v}\nonumber 
\end{eqnarray}
Then project the axis 3 to have zero values using $P^{\prime}$ matrix
\begin{eqnarray*}
P^{\prime} & = & \left(\begin{array}{ccc}
1 & 0 & 0\\
0 & 1 & 0\\
0 & 0 & 0
\end{array}\right)
\end{eqnarray*}
Then rotate to the original axis system using $R_{\boldsymbol{u}}^{\dagger}$.
The entire operation can be summed up as
\begin{eqnarray}
P & = & R_{\boldsymbol{u}}^{\dagger}P^{\prime}R_{\boldsymbol{u}}\label{eq:constrained-projection-3D}\\
\boldsymbol{r}^{\prime} & = & P\boldsymbol{r}\nonumber 
\end{eqnarray}
The coordinates $\boldsymbol{r}^{\prime}$defined using the projection
will satisfy $\boldsymbol{v}\cdot\boldsymbol{r}^{\prime}=0$ for arbitrary
$\boldsymbol{r}$. Applying this projection operator from eq. (\ref{eq:constrained-projection-3D})
on most general group element from eq. (\ref{eq:YPR}) will produce
a group homomorphic to the constrained sub-group required. The Lie-algebra
elements of this group can now be found using eq. (\ref{eq:differential-def-Lie-algebra}).
The matrices obtained using this method are in general not orthogonal,
but if their linear combinations are considered, a subspace of them
will be found to be orthogonal. These elements of subspace will be
\begin{eqnarray*}
J_{\hat{\boldsymbol{v}}} & = & \hat{\boldsymbol{v}}\cdot\boldsymbol{J}\\
\boldsymbol{J} & = & \left(J_{1},J_{2},J_{3}\right)
\end{eqnarray*}
and the new rotation matrix
\begin{eqnarray*}
R_{\boldsymbol{v}} & = & \text{exp}\left(J_{\hat{\boldsymbol{v}}}\theta\right)
\end{eqnarray*}
 is the required element of the constrained sub-group of $SO\left(3\right)$.
These generators can also be found by operating projection operator
on either sides of generators of transformation $\boldsymbol{J}$as
follows:

\begin{eqnarray*}
J_{\hat{v}} & = & P\boldsymbol{J}P=\underset{\theta_{i}\rightarrow0}{\text{lim}}\frac{\partial}{\partial\theta_{i}}P\text{exp}\left(-\boldsymbol{J}\cdot\boldsymbol{\theta}\right)P
\end{eqnarray*}

In order to construct similar projection operators for more than 3
dimensions, which are required for solving many-body problems, one
need to find the linearly independent triplets of vectors which can
be expressed in zero sum (difference) form. These sums and the orthogonality
condition (or operating the projection matrix on either sides) can
be used to find the rotation matrices equivalent to the one in eq.
(\ref{eq:Rot-preProj}), which after projection like in eq. (\ref{eq:constrained-projection-3D})
would give the required projection operators for polyhedron constraints.

Another way to obtain projection operator is by finding coefficient
matrices $\left(A_{m\times n}\right)$ of the linear equations in
$n$ variables, that are typically $m\times n$ $\left(m<n\right)$
dimenstional. The projection matric in that case is given by:

\begin{eqnarray*}
P_{\perp} & = & I_{n\times n}-\left(A^{-1}\right)_{n\times m}A_{m\times n}
\end{eqnarray*}

where $\left(A^{-1}\right)_{n\times m}$ is pseudo-inverse of matrix
$A_{m\times n}$.

\section*{Acknowledgments}
These ideas have been developing in author's mind since November 2013. Author thanks his teachers from Indian Institute of Technology Kanpur, Tata Institute of Fundamental Research, and Case Western Reserve University, for teaching the essential concepts of physics.

\bibliographystyle{unsrt}  
\bibliography{manuscript}

\end{document}